\documentclass[prd,aps,preprintnumbers,showpacs,10pt]{revtex4}
\usepackage{epsfig}
\renewcommand{\arraystretch}{1.8}

\def\g{\gamma}

\def\p{\pi}      
\def\q{\theta}   
\def\r{\rho}

\newcommand{\be}{\begin{equation}}
\newcommand{\ee}{\end{equation}}
\newcommand{\bea}{\begin{eqnarray}}
\newcommand{\eea}{\end{eqnarray}}
\newcommand{\nn}{\nonumber}
\newcommand{\dd}{\displaystyle}

\def\slash#1{\setbox0=\hbox{$#1$}#1\hskip-\wd0\dimen0=5pt\advance
\dimen0 by-\ht0\advance\dimen0 by\dp0\lower0.5\dimen0\hbox
to\wd0{\hss\sl/\/\hss}} 
\begin{document}
\begin{titlepage}

\preprint{BARI-TH 531/06}

\title{Non leptonic $B$ decays to axial-vector mesons and factorization}

\author{\textbf {V. Laporta}}\author{\textbf{G. Nardulli}}
\affiliation{Dipartimento di Fisica dell'Universit{\`a} di Bari, Italy\\
Istituto Nazionale di Fisica Nucleare, Sezione di Bari, Italy}
\author{\textbf{T. N. Pham}}
\affiliation{Centre de Physique Th{\'e}orique,\\
Centre National de la Recherche Scientifique, UMR 7644, \\
{\'E}cole Polytechnique, 91128 Palaiseau Cedex, France}

\begin{abstract}We present an
analysis of two-body $B$ decays with a pseudoscalar ($P$) and an
axial-vector meson ($\cal A$) in the final state using
factorization. We employ  as inputs a limited number of experimental
data, i.e.
 results for the  $B\to K_1\g~,$ and $B\to~ K^*\g$ radiative
decays and the branching ratios for $B\to \pi\rho,~\pi
K^*,~K\rho,~K\pi$ non leptonic decays.  In this way we are able to
compare our predictions with recent data from the Belle and BABAR
collaborations on $B\to a_1\pi$ and make predictions on several
other $B\to P\cal A$ decay channels, which might be used as a
guide for experimental researches and as tests of factorization.
\end{abstract}


\maketitle
\end{titlepage}
\setcounter{page}{1}
\newpage
\section{Introduction}
Recent observations of the decays $B^0\to a_1^\pm(1260)\pi^\mp$
from the Belle \cite{Abe:2005rf} and the BABAR Collaborations
\cite{Aubert:2006dd} offer the possibility of new investigations
for two-body decay channels of the $B$ meson with an axial-vector
meson in the final state. The BABAR result \be {\cal B}(B^0\to
a_1^\pm(1260)\pi^\mp)\times {\cal B}(a_1^\pm(1260)\to
\pi^\pm\pi^\pm\pi^\mp) =(16.6\pm 1.9\pm 1.5)\times 10^{-6}\ee
translates into \be {\cal B}(B^0\to a_1^\pm(1260)\pi^\mp)
=(33.2\pm 3.8\pm 3.0)\times 10^{-6}\label{1}\ee assuming
\cite{Aubert:2006dd} that $a_1$ only decays into three pions and
an equal yield for $a_1^\pm(1260)\to \pi^\pm\pi^\pm\pi^\mp $and
for $a_1^\pm(1260)\to \pi^\pm\pi^0\pi^0$. On the other hand the
Belle measurement gives: \be
      {\cal B}(B^0\to  a_1^{\pm}(1260)\pi^\mp ) =
      (48.6 \pm 4.1\pm 3.9)\times 10^{-6}\ ,\label{belle}\ee
      with an average of the two experiments
      \be {\cal B}(B^0\to  a_1^{\pm}(1260)\pi^\mp ) =(40.9\pm 7.6)\times 10^{-6}\ .\ee
      In a recent paper
\cite{Nardulli:2005fn}, hereafter referred to as I, two of us have
discussed other two-body non leptonic decays of the $B$ meson with
an axial vector meson in the final state and proposed some simple
tests of factorization for them. The  analysis of I was stimulated
by the experimental results ${\cal B}(B^+\to K_1^+(1270)\gamma)=
(4.28\pm0.94\pm0.43)\times 10^{-5}$~ and ${\cal B}(B^+\to
K_1^+(1400)\gamma)< 1.44\times 10^{-5}~({\rm at ~90\% ~C.L.})$
from the Belle collaboration \cite{Abe:2004kr}. These numerical
results are comparable with data for the analogous channels with a
vector meson in the final state: $ {\cal B}( B^{+}\, \to K^{\ast
+}\gamma)=(4.18\pm 0.31)\times 10^{-5}$ and  $ {\cal B}(B^0\, \to
K^{\ast 0} \gamma)=(4.17\pm 0.23)\times 10^{-5}$ ({\rm averages of
\cite{Coan:1999kh,Aubert:2001me,Nakao:2004th}}). Therefore they
suggest an approximate equality between the form factors for $B\to
$ vector and $B\to$ axial-vector transitions
\cite{Nardulli:2005fn}. Using this simple observation, in I we
have proposed several tests of factorization for the $B$ decay
channels with a strange axial-vector meson in the final state. In
this paper we wish to reconsider this approach and to extend it to
other decay channels with strange particles in the final state as
well as final states with no strange particles. In particular we
wish to compare theoretical expectations with the  BABAR and Belle
results (\ref{1}) and (\ref{belle}), and to give predictions for
several similar decay channels that have not been examined  yet
theoretically, but might be studied by the BABAR and other
experimental collaborations. The study of two-body charmless $B$
decays with a light pseudo-scalar meson and an axial meson in the
final states, besides providing us with information on the $B\to
a_{1}$ and $B\to K_{1}$ transition form factors, could also tell
us about the dynamics of these decays modes. Unlike the $B\to
K\rho$ decays which is much suppressed because of the destructive
interference of the $O_{4}$ and $O_{6}$ matrix elements, the
decays $B\to a_{1} K$ could have a large branching ratio ({\it
BR}) , since the interference term becomes constructive and
enhances the decay rates as in  the $B\to K\pi$ decay. Therefore a
large {\it BR} similar to the {\it BR} for $B\to K\pi$ would be a
confirmation of a large $B\to a_{1}$ transition form factors and
the penguin dominance of this decay.

Our approach is based on the idea that factorization, together with
 experimental data for the {\it BRs} of the decays
 $B\to K^*\p\,,K\p\,,\,\r\p,\,\r K$, can provide enough information to predict
nonleptonic $B$-decays with one axial-vector meson in the final
state. It is known that factorization  holds only approximately
and in some cases its predictions are at odds with experiment. In
the last few years factorization has been proved to be a rigorous
prediction of QCD in the infinite quark mass limit
\cite{BBNS:1999,Beneke:2000ry,Beneke:2003zv} and the naive
factorization scheme has evolved into a more precise approach,
using effective theories and an expansion in $\Lambda_{QCD}/m_b$.
In this context it should be noted however that for $B$ decays
into two light hadrons a proof at all orders is still missing;
moreover, for  charmless $B$  decays with a strange light hadron
in the final state, the tree-level $O_1,O_2$ operators are CKM
suppressed compared with the $O_4, O_6$ matrix elements. This
gives a numerically important contribution to the
penguin-dominated decays since $O_6$ matrix elements are
chirally-enhanced in  naive and in QCD factorization, although
power-suppressed by inverse powers of $m_b$ \cite{Beneke:2003zv}
if annihilation terms are neglected.

To our knowledge there is currently no extensive study of
charmless $B$ decays with a final axial vector meson based on
factorization. Therefore we feel it can be useful to collect
predictions on these channels using the simple naive factorization
approach, though we are conscious that these results should be
interpreted with some care and used more as guidelines for
experiment than as absolute predictions. The advantage over
previous calculations of some related decay channels using
factorization, e.g. \cite{Bauer:1986bm,Ali:1998eb,Deandrea:1993},
is the fact that we do not use predictions from theoretical models
for the form factors. Therefore any discrepancy that  might be
found between our predictions and future data would point  to a
breakdown  of naive factorization and suggest more refined
treatments.

 The plan of the paper is as follows. After
a review of the approach in section \ref{sec:2}, we apply this
method to get predictions for $B\to K_1\pi$ in section \ref{k1p},
for $B\to a_1 K$ and $B\to b_1 K$ in section \ref{sec:4} and for
$B\to a_1\pi\,,\, b_1\pi$ in section \ref{sec:5}. Section
\ref{sec:6} contains our conclusions, while in Appendix A we have
collected some relevant formulae used in the main text.

\section{Method and definitions\label{sec:2}}Let us
start with some relations among the various form factors of the
$V-A$ currents that will be used below. We use definitions of form
factors as listed in Appendix A.  The main idea of I
 was to use ratios of {\it BRs} to
deduce ratios of form factors and, subsequently,  to use this piece
of information to predict decays of the $B$-meson into final states
with an axial-vector meson. To this effect we will need below the
ratios $\dd\frac{A_0^{B\to V}}{F_{0,1}^{B\to P}}$ ,
$\dd\frac{F_0^{B\to K}}{F_0^{B\to\p}}$ and $\dd\frac{V_0^{B\to{\cal
A}}}{F_0^{B\to V}}$ where $P,V$ and ${\cal A}$ refer to
pseudoscalar, vector and axial-vector meson. We will determine these
quantities by factorization and using experimental data.

As a matter of fact factorization predicts the following results
\bea \frac{{\cal B}(B^{+}\to K^0\,\r^+)}{{\cal B}(B^{+}\to
K^0\,\p^+)}&\simeq&\frac{4}{m_B^2}\left|\frac{W_8} {W_1}\right|^2
\left(\frac{A_0^{B\to\r}(m_K^2)}{F_0^{B\to\p}(m_K^2)}\right)^2
\frac{q_\r^3}{q_\p}\label{KR}
\\
\frac{{\cal B}(B^{0}\to K^+\,\r^-)}{{\cal B}(B^{0}\to
K^+\p^-)}&\simeq&\frac{4}{m_B^2}\left|\frac{W_6} {W_2}\right|^2
\left(\frac{A_0^{B\to\r}(m_K^2)}{F_0^{B\to\p}(m_K^2)}\right)^2
\frac{q_\r^3}{q_\p}\label{A0rho}
\\
\frac{{\cal B}(B^{0}\to K^0\,\r^0)}{{\cal B}(B^{0}\to
K^0\,\p^0)}&\simeq&\frac{4}{m_B^2}\left|\frac{W_8\frac{\,A_0^{B\to\r}(m_K^2)}{F_0^{B\to
\p}(m_K^2)}+W_7\frac{\,f_\r}{f_K}\,\frac{F_1^{B\to
K}(m_\r^2)}{F_0^{B\to
\p}(m_K^2)}}{W_1+W_3\,\frac{f_\p}{f_K}\,\frac{F_0^{B\to
K}(m_\p^2)}{F_0^{B\to \p}(m_K^2)}}\right|^2
\frac{q_\r^3}{q_\p}\label{F0kpi}
\\
\frac{{\cal B}(B^{+}\to K^+\,\r^0)}{{\cal B}(B^{+}\to
K^+\,\p^0)}&\simeq&\frac{4}{m_B^2}\left|\frac{W_6\frac{\,A_0^{B\to\r}(m_K^2)}{F_0^{B\to
\p}(m_K^2)}+W_7\frac{\,f_\r}{f_K}\,\frac{F_1^{B\to
K}(m_\r^2)}{F_0^{B\to
\p}(m_K^2)}}{W_2+W_3\,\frac{f_\p}{f_K}\,\frac{F_0^{B\to
K}(m_\p^2)}{F_0^{B\to \p}(m_K^2)}}\right|^2 \frac{p_\r^3}{p_\p}
\\
\frac{{\cal B}(B^{+}\to K^{*0}\,\p^+)}{{\cal B}(B^{+}\to
K^0\,\p^+)}&\simeq&\frac{4}{m_B^2}\left|\frac{W_4} {W_1}\right|^2
\left(\frac{f_{K^*}}{f_K}\right)^2 \frac{q_{K^*}^3}{q_K}
\\
\frac{{\cal B}(B^{0}\to K^{*+}\,\p^-)}{{\cal B}(B^{0}\to
K^+\p^-)}&\simeq&\frac{4}{m_B^2}\left|\frac{W_5}
{W_2}\right|^2\left(\frac{f_{K^*}}{f_K}\right)^2
\frac{q_{K^*}^3}{q_K}
\\
\frac{{\cal B}(B^{0}\to K^{*0}\,\p^0)}{{\cal B}(B^{0}\to
K^0\,\p^0)}&\simeq&\frac{4}{m_B^2}\left|\frac{W_3\frac{\,f_\p}{f_{K^*}}\,\frac{A_0^{B\to
K^*}(m_\p^2)}{F_1^{B\to
\p}(m_{K^*}^2)}+W_4}{W_3\,\frac{f_\p}{f_{K^*}}\,\frac{F_0^{B\to
K}(m_\p^2)}{F_1^{B\to
\p}(m_{K^*}^2)}+W_1\frac{f_K}{f_{K^*}}}\right|^2
\frac{q_{K^*}^3}{q_K}
\\
\frac{{\cal B}(B^{+}\to K^{*+}\,\p^0)}{{\cal B}(B^{+}\to
K^+\,\p^0)}&\simeq&\frac{4}{m_B^2}\left|\frac{W_3\frac{\,f_\p}{f_{K^*}}\,\frac{A_0^{B\to
K^*}(m_\p^2)}{F_1^{B\to
\p}(m_{K^*}^2)}+W_5}{W_3\,\frac{f_\p}{f_{K^*}}\,\frac{F_0^{B\to
K}(m_\p^2)}{F_1^{B\to
\p}(m_{K^*}^2)}+W_2\frac{f_K}{f_{K^*}}}\right|^2
\frac{q_{K^*}^3}{q_K}\label{eq:kstarpi}\ .\eea We have indicated
the squared meson mass in the argument of the form factors to keep
track of the factorization procedure, but in the numerical
computations all form factors are evaluated at $q^2=0$, which
should be a rather good approximation;
 ${q_{\r}}$, ${q_\p}$, ${q_{K^*}}$ and ${q_K}$ are momenta
 in the $B$ rest frame and
 $W_i$ are combinations of Wilson coefficients and CKM matrix elements
that can be found in the upper part of the Table reported in
Appendix A. Using as inputs the experimental ratios for
$\dd\frac{{\cal B}(B^{0}\to K^+\,\r^-)}{{\cal B}(B^{0}\to K^+\p^-)}$,
$\dd\frac{{\cal B}(B^{0}\to K^0\,\r^0)}{{\cal B}(B^{0}\to K^0\,\p^0)}$, and
$\dd\frac{{\cal B}(B^{0}\to K^{*0}\,\p^0)}{{\cal B}(B^{0}\to K^0\,\p^0)}$,
we have computed the entries in the first columns in Table
\ref{tab:fff} (the last column is obtained by the ratio of the first
two data).

\begin{table}[ht!]
\caption{\label{tab:fff} {\small Ratio of form factors involving $B$
decays to negative parity mesons. The results of the present paper
can be compared to the findings obtained by the Bauer-Stech-Model
(BSW), the Heavy Meson Effective Lagrangian (HMEL), Light-Cone Sum
Rules (LCSR) and the Covariant Light Front Approach (CLFA).}}
\begin{center}
\begin{tabular}{|c|c|c|c|c|}
\hline
~~~~~~~~~~~~~~~~~~~~~~~~&~~~$\dd\frac{A_0^{B\to\r}(0)}{F_1^{B\to\p}(0)}$~~~&~~~
$\dd\frac{F_1^{B\to
K}(0)}{F_1^{B\to\p}(0)}$~~~&~~~$\dd\frac{A_0^{B\to
K^*}(0)}{F_1^{B\to\p}(0)}$~~~&~~~$\dd\frac{A_0^{B\to\r}(0)}{F_1^{B\to
K}(0)}$~~~
\\
\hline BSW \cite{Bauer:1986bm}&0.84&1.15&0.97&0.73
\\
\hline HMEL
\cite{Deandrea:1993}&$0.45\pm0.56$&$0.92\pm0.32$&$0.38\pm0.46$&$0.49\pm0.60$
\\
\hline LCSR \cite{Ball:2003rd}&1.15&1.30&1.38&0.88
\\
\hline CLFA \cite{Cheng:2004yj}&1.12&1.40&1.24&0.80
\\
\hline This work&1.63&1.56&1.98&1.02
\\
\hline
\end{tabular}
\end{center}
\end{table}
We also present a comparison  with other theoretical approaches. We
notice that our predictions for the ratios are in general higher
than other methods. The Light Cone Sum Rules (LCSR) results of Refs.
\cite{Ball:2003rd} are however the less distant from ours.

We can now use these data to compute the remaining {\it BRs} in
Eqns.(\ref{KR})-(\ref{eq:kstarpi}). The results are reported in
Table \ref{BR} and  can be considered as a consistence test for the
method to be used in the subsequent Sections. In particular we note
 that the ratios $\dd\frac{{\cal B}(B^{+}\to K^+\,\r^0)}{{\cal
B}(B^{+}\to K^+\,\p^0)}$ and $\dd\frac{{\cal B}(B^{+}\to
K^{*+}\,\p^0)}{{\cal B}(B^{+}\to K^+\,\p^0)}$ agree  with the
experimental results. It can be also noticed that $\dd\frac{{\cal
B}(B^{+}\to K^{*0}\,\p^+)}{{\cal B}(B^{+}\to K^0\,\p^+)}$ and
$\dd\frac{{\cal B}(B^{0}\to K^{*+}\,\p^-)}{{\cal B}(B^{0}\to K^+\p^-)}$ in
the present approximation are completely independent of form
factors.

\begin{table}[ht!]
\caption{\label{BR} {\small  Ratios of Branching Ratios and their
comparison with experiment.}}
\begin{center}
 \begin{tabular}{|c|c|c||c|c|c|}
\hline ~~~~~~~{\rm Ratios}~~~~~~~&~~~~~~~Th.~~~~~~~&~~~~~~~Exp.
\cite{Group(HFAG):2005rb},\cite{Aubert:2007mb}~~~~~~~ &~~~~~~~~{\rm
Ratios}~~~~~~~~&~~~~~~~Th.~~~~~~~&~~~~~~~Exp.
\cite{Group(HFAG):2005rb}~~~~~~~
\\
\hline $\dd\frac{{\cal B}(B^{0}\to K^+\,\r^-)}{{\cal B}(B^{0}\to
K^+\p^-)}$&input&$0.54\pm0.11$&$\dd\frac{{\cal B}(B^{0}\to
K^{*+}\,\p^-)}{{\cal B}(B^{0}\to K^+\p^-)}$&0.55&$0.69\pm0.13$
\\
\hline $\dd\frac{{\cal B}(B^{0}\to K^0\,\r^0)}{{\cal B}(B^{0}\to
K^0\,\p^0)}$&input&$0.44\pm0.17$&$\dd\frac{{\cal B}(B^0\to
K^{*0}\,\p^0)}{{\cal B}(B^{0}\to K^0\,\p^0)}$&input&$0.14\pm0.08$
\\
\hline $\dd\frac{{\cal B}(B^{+}\to K^+\,\r^0)}{{\cal B}(B^{+}\to
K^+\,\p^0)}$&0.42&$0.42\pm0.10$&$\dd\frac{{\cal B}(B^+\to
K^{*+}\,\p^0)}{{\cal B}(B^{+}\to K^+\,\p^0)}$&0.71&$0.57\pm0.22$
\\
\hline $\dd\frac{{\cal B}(B^{+}\to K^0\,\r^+)}{{\cal B}(B^{+}\to
K^0\,\p^+)}$&0.018&$0.35\pm 0.07$& $\dd\frac{{\cal B}(B^{+}\to
K^{*0}\,\p^+)}{{\cal B}(B^{+}\to K^0\,\p^+)}$&0.46&$0.40\pm0.07$
\\
\hline
\end{tabular}
\end{center}
\end{table}
We notice a discrepancy between theoretical expectations  based on
factorization and experimental results for the ratio $\dd\frac{{\cal
B}(B^{+}\to K^0\,\r^+)}{{\cal B}(B^{+}\to K^0\,\p^+)}$. The data
come from the experimental value $\dd{\cal B}(B^{+}\to
K^0\,\p^+)=(23.1\pm1.0)\times10^{-6}$ \cite{Group(HFAG):2005rb} and
the new finding from BABAR \cite{Aubert:2007mb} \be {\cal
B}(B^{+}\to K^0\,\r^+)= \left(8.0^{+1.4}_{-1.3}\pm0.5\right)\times
10^{-6}\ .\label{BRpred}\ee It seems to show a significant effect
beyond naive factorization, e.g. one due to final state
interactions, see the discussion below.

In order to apply this method to the decays with an axial-vector
meson in the final state we need information on the corresponding
form factors, whose definition is in the Appendix.
 In I we assumed that the
effect of substituting $K^*$ with $K_1$ is identical in the
radiative and in the non-leptonic decay, in other words that each
form factor for the $B\to K_1$ transition is given by the
corresponding form factor for $B\to K^*$ multiplied by the same
factor $y$, once the change of parity between the two strange mesons
and the kinematical factors are taken into account. For our purposes
only the form factor $V_0$ (see the Appendix) for the transition
$B\to$ axial-vector meson is relevant. Using the above-mentioned
assumption we get \be V_0^{B\to K_1(1270)}(q^2)=h\ A_0^{B\to
K^*}(q^2) ~,~~~~~ V_0^{B\to K_1(1400)}(q^2)=k\ A_0^{B\to K^*}(q^2)
\label{eqV0}\ee with \be \left(\begin{array}{c}
          h \\
          k
        \end{array}\right) =
\frac{m_{K^*}}{m_{K_1}}\frac{m_B+m_{K_1}-(m_B-m_{K_1})z}{m_B+m_{K^*}-(m_B-m_{K^*})z}
\,\left(\begin{array}{c}
          y \\
          y^\prime
        \end{array}\right)\,,\label{hk}\ee
where $y$ and $y'$ are defined in the Appendix, while the factor
$z$ is defined as \be
z=\frac{A_2^{B\to\rho}(0)}{A_1^{B\to\rho}(0)}\approx
\frac{A_2^{B\to K^*}(0)}{A_1^{B\to K^*}(0)}\,.\ee We take the
value $z=0.93$, intermediate between the value predicted by light
cone sum rules \cite{Ball:2003rd} ($z=0.9$) and that given by the
BSW model \cite{Bauer:1986bm} ($z=0.95$). In the following we will
need also of the ratio $\frac{\dd V_0^{B\to {\cal A}_1}}{\dd
A_0^{B\to\r}}$ (with ${\cal A}_1=a_1$ or $b_1$); we can predict it
from the previous result:
\begin{eqnarray} \frac{V_0^{B\to a_1}(0)}{A_0^{B\to\r}(0)}
&\approx&\frac{V_0^{B\to K_{1A}}(0)}{A_0^{B\to
K^*}(0)}~=~h\sin\q+k\cos\q\label{eq.V0A01}\ ,
\\
\frac{V_0^{B\to b_1}(0)}{A_0^{B\to\r}(0)}&\approx&\frac{V_0^{B\to
K_{1B}}(0)}{A_0^{B\to K^*}(0)}~=~h\cos\q-k\sin\q\ .\label{eq.V0A02}
\end{eqnarray}
In previous equations we assume that the ratios satisfy $SU(3)$
flavor symmetry to a good approximation since $SU(3)$ breaking
terms tend to cancel out in the ratio, see e.g. \cite{Ali:1998eb}.
In equations (\ref{eq.V0A01}) and (\ref{eq.V0A02}) $\theta$ is the
mixing angle between the octets $^3P_1$ and $^1P_1$ from which the
states $K_1(1270)$ and $K_1(1400)$ result. To the former octet
belong $a_1$ and the unmixed strange state $K_{1A}$; to the latter
$b_1$ and $K_{1B}$, see the Appendix A for further details. The
mixing scheme we adopt here is analogous to that based on the
conventional quark model of Ref. \cite{Suzuki:1993yc};  $\theta$
is the mixing angle between two strange P wave axial meson;
therefore, differently from, e.g. $\eta-\eta^\prime$ mixing, it
should not be affected by gluonium contributions.  The
phenomenological analysis
\cite{Suzuki:1993yc,Cheng:2004yj,Nardulli:2005fn} gives as
possible results $\q=32^0$ or $58^0$.
 In Table \ref{tab:fff2} we
report our predictions for both values and a comparison with the
result of Ref. \cite{Cheng:2004yj}.

\begin{table}[h!]
\caption{\label{tab:fff2} {\small  Ratio of form factors for $B$
decays to axial-vector mesons.}}
\begin{center}
\begin{tabular}{|c|c|c|}
\hline

~~~~~~~~~~~~~~~~~~~~~~~~~~~~~~~~&~~~$\dd\frac{V_0^{B\to
K_{1A}}(0)}{A_0^{B\to K^*}(0)}$~~~&~~~$\dd\frac{V_0^{B\to
K_{1B}}(0)}{A_0^{B\to K^*}(0)}$~~~
\\
\hline This work ($\q=32^o$)&0.64&0.78
\\
This work ($\q=58^o$)&1.02&0.26
\\
\hline CLFA \cite{Cheng:2004yj}&0.45&1.32
\\
\hline
\end{tabular}
\end{center}
\end{table}

\section{$B\to K_1\pi$\label{k1p}} These channels
were already considered in I and we report them for completeness. If
$q_{K_1}$ and $q_{K^*}$ are respectively the c.m. momenta of ${K_1}$
and ${K^*}$ in the reactions $B\to K_1 \pi$ and $B\to K^{*} \pi$,
one gets, using factorization: \be\frac{{\cal B}(B^+\to K^0_1
\pi^+)_{\rm fact.}}{{\cal B}(B^+\to K^{*\,0} \pi^+)_{\rm fact.}}\,
=\, \frac{{\cal B}(B^0\to K^+_1 \pi^-)_{\rm fact.}}{{\cal B}(B^0\to
K^{*\,+} \pi^-)_{\rm fact.}}\,= \,
\left(\frac{q_{K_1}}{q_{K^*}}\right)^3\,\left(\frac { F_1^ {B\to\pi}
(m^2_{K_1}) \,f_{K_1}  } { F_1^ {B\to\pi} (m^2_{K^*}) f_{K^*}
}\right)^2\label{pred0}\ ,\ee where the subscript means that we
consider only factorizable  contributions. Therefore, using
$f_{K_1}$ from $\tau$ decays (see the Appendix) one can predict
${\cal B}(B^+\to K^0_1 \pi^+)$ and ${\cal B}(B^0\to K^+_1 \pi^-)$
for both $K_1(1270)$ and $K_1 (1400)$ from the knowledge of ${\cal
B}(B^+\to K^{*\,0} \pi^+)$ and ${\cal B}(B^0\to K^{*\,+} \pi^-)$
\cite{Eidelman:2004wy}.

\begin{table}[ht!]
\caption{\label{tab:1} {\small  Theoretical branching ratios  for
$B$ decays into a strange axial-vector meson and a pion. Units
$10^{-6}$.}}
\begin{footnotesize}
\begin{center}
\begin{tabular}{|c|c|c|}
\hline

~~~~~~~~~~~{\rm Process}~~~~~~~~~~~  & ~~~~~~~~~~~${\cal B}$~({\rm
Th.})~~~~~ & ~~~
~~${\cal B}~ ({\rm Exp.})$~ \cite{Eidelman:2004wy} \\

\hline\hline

$  B^{+}\, \to \pi^{+}~K_1^0(1270)$ & 5.8 & ==\\
\hline

$  B^{+}\, \to \pi^0~K_1^+(1270)$ & 4.9 & ==\\
\hline

$  B^{0}\, \to \pi^0~K_1^0(1270)$ & 0.4 & ==\\
\hline

$  B^{0}\, \to \pi^{-}~K_1^+(1270)$ &  7.6 & ==\\
\hline\hline

$  B^{+}\, \to \pi^{+}~K_1^0(1400)$ &  3.0 &$<\,260$ \\
\hline

$B^{+}\, \to \pi^{0}~K_1^+(1400)$ & \begin{tabular}{c}1.0~~
($\theta=32^0$)\\1.4~~($\theta=58^0$)\end{tabular} &==\\
\hline

$B^{0}\, \to \pi^{0}~K_1^0(1400)$ & \begin{tabular}{c}
3.0~~($\theta=32^0$)\\1.7~~($\theta=58^0$)
\end{tabular} &==\\
\hline

$  B^{0}\, \to \pi^{-}~K_1^+(1400)$ &  4.0&$<\,1100$\\

\hline
\end{tabular}
\end{center}
\end{footnotesize}
\end{table}
The  reactions with a $\pi^0$ in the final state: $B^+\to
K^+_1\pi^0$,  $B^0\to K^0_1\pi^0$ involve three form factors $F_1$,
$A_0$ and $V_0$ and different combinations of Wilson coefficients
and CKM matrix elements. One gets ($s=h,~k$, see Eq. (\ref{hk}), for
$K_1(1270)$ and $K_1(1400)$ respectively): \bea \frac{{\cal
B}(B^{+}\to K_1^+\,\p^0)_{\text{ fact.}}}{{\cal B}(B^{+}\to
K^{*+}\,\p^0)_{\text{
fact.}}}&=&\left(\dd\frac{q_{K_1}}{q_{K^*}}\right)^3\,\left|\frac{\dd
\frac{W_5}{W_3}\frac{f_{K_1}}{f_\p}\frac{\dd
F_1^{B\to\p}(m^2_{K_1})}{A_0^{B\to K^*}(m^2_\p)}\,+\,s}
{\frac{W_5}{W_3}\frac{f_{K^*}}{f_\p}\frac{\dd
F_1^{B\to\p}(m^2_{K^*})}{A_0^{B\to K^*}(m^2_\p)}+1}\right|^2
\\
\frac{{\cal B}(B^{0}\to K_1^0\,\p^0)_{\text{ fact.}}}{{\cal B}(B^{0}\to
K^{*0}\,\p^0)_{\text{
fact.}}}&=&\left(\dd\frac{q_{K_1}}{q_{K^*}}\right)^3\,\left|\frac{\dd
\frac{W_4}{W_3}\frac{f_{K_1}}{f_\p}\frac{\dd
F_1^{B\to\p}(m^2_{K_1})}{A_0^{B\to K^*}(m^2_\p)}\,+\,s}
{\frac{W_4}{W_3}\frac{f_{K^*}}{f_\p}\frac{\dd
F_1^{B\to\p}(m^2_{K^*})}{A_0^{B\to K^*}(m^2_\p)}+1}\right|^2\,; \eea
where $W_i$ are listed in Appendix. The result of this analysis is
in Table \ref{tab:1}. For the form factors ratios, that we have
considered at $q^2=0$, we have used the values in Table
\ref{tab:fff}.

\section{$B\to {\cal A}_1 K$\label{sec:4}}
In this section we consider the decays $B\to a_1 K\ ,~b_1 K$. Also
in this case we have some clear predictions based on factorization
for the decays with a charged axial-vector meson in the final state
\cite{Nardulli:2005fn}:
 \bea
 \frac{{\cal B}(B^+\to {\cal A}_1^+ K^0)_{\rm fact.}}{{\cal
B}(B^+\to \rho^+ K^{0})_{\rm fact.}} &= &\left(\frac{q_{{\cal
A}_1}}{q_{\rho}}\right)^3\,\left|\frac{W_1}{W_8}\frac{V_0^{B\to
{\cal A}_1}(m_K^2)}{A_0^{B\to\r}(m_K^2)}\right|^2\,,\label{a1}
\\
&& \cr\frac{{\cal B}(B^0\to {\cal A}_1^- K^+)_{\rm fact.}}{{\cal
B}(B^0\to \rho^- K^{+})_{\rm fact.}} &=&\left(\frac{q_{{\cal
A}_1}}{q_{\rho}}\right)^3\, \left|\frac{W_2}{W_6}\frac{V_0^{B\to
{\cal A}_1}(m_K^2)}{A_0^{B\to\r}(m_K^2)}\right|^2\label{a3} \,,\eea
where $\dd\frac{V_0^{B\to {\cal A}_1}(m_K^2)}{A_0^{B\to\r}(m_K^2)}$
is given by Eq.(\ref{eq.V0A01}) or Eq. (\ref{eq.V0A02}) for ${\cal
A}_1=a_1$ or $b_1$ respectively.

Similar predictions can be given also for the channels with a
neutral axial-vector meson in the final state, ie the decay channels
$B^+\to a_1^0\,K^+$, $B^+\to b_1^0\, K^+$, $B^0\to a_1^0\, K^0$ and
$B^0\to b_1^0\, K^0$, though the corresponding formulae are more
involved. In fact we have\bea
 \frac{{\cal B}(B^+\to {\cal A}_1^0\, K^+)_{\rm fact.}}{{\cal
B}(B^+\to \rho^0\, K^{+})_{\rm fact.}} &= &\left(\frac{q_{{\cal
A}_1}}{q_{\rho}}\right)^3\,\left|\frac{\frac{V_0^{B\to {\cal
A}_1}(m_K^2)}{A_0^{B\to\r}(m_K^2)}\,W_2\,+W_3\,\frac{\dd f_{{\cal
A}_1}}{f_K}\frac{\dd F_1^{B\to K}(m^2_{{\cal
A}_1})}{A_0^{B\to\r}(m^2_K)}}{W_6\,+\,W_7\,\frac{\dd
f_\r}{f_K}\frac{\dd F_1^{B\to
K}(m^2_\r)}{A_0^{B\to\r}(m^2_K)}}\right|^2\ ,\\ && \cr &&\cr
\frac{{\cal B}(B^0\to {\cal A}_1^0\, K^0)_{\rm fact.}}{{\cal
B}(B^0\to \rho^0\, K^{0})_{\rm fact.}} &=&\left(\frac{q_{{\cal
A}_1}}{q_{\rho}}\right)^3\,\left|\frac{ \frac{V_0^{B\to {\cal
A}_1}(m_K^2)}{A_0^{B\to\r}(m_K^2)}\,W_1\,+W_3\,\frac{\dd f_{{\cal
A}_1}}{f_K}\frac{\dd F_1^{B\to K}(m^2_{{\cal
A}_1})}{A_0^{B\to\r}(m^2_K)}}{W_8\,+\,W_7\,\frac{\dd
f_\r}{f_K}\frac{\dd F_1^{B\to
K}(m^2_\r)}{A_0^{B\to\r}(m^2_K)}}\right|^2\label{a3bis}\,;\eea the
ratio $\frac{F_1^{B\to K}}{A_0^{B\to\r}}$ can be computed at
$q^2=0$, as reported in Table \ref{tab:fff}; the coefficients $W_i$
are in the Appendix A. The results obtained are reported in Table
\ref{tab:2}.  We have used the experimental {\it BR} for $B\to
\rho^- K^+$ as given by the HFAG group
  \cite{Group(HFAG):2005rb}. For the channels $B^{+}\, \to K^0~a_1^+$
  and $B^{+}\, \to K^0~b_1^+$ we have reported two values. The former is
  based
on the  result (\ref{BRpred}) (which supersedes an old upper limit
 \cite{Eidelman:2004wy}) for the {\it BR} of the
  channel $B^+\to \rho^+ K^0$; the latter (given in parentheses)
  is based the experimental result \cite{Group(HFAG):2005rb} for the {\it BR} of the channel
  $B^+\to \pi^+ K^0$, according to the discussion below.

\begin{table}[ht!]
\caption{\label{tab:2} {\small  Theoretical branching ratios  for
$B$ decays into one nonstrange axial-vector meson and a kaon for two
different values of the mixing angle. Units $10^{-5}$}. Predictions
use $B\to K\rho$ decay rates for all entries except those given in
parentheses, where the experimental BR for $B\to K^0\pi^+$ is used.}
\begin{footnotesize}
\begin{center}
\begin{tabular}{|c|c|c|}
\hline ~~~~~~{\rm Process}~~~~~~  & ~~~${\cal B}\,\,
(\theta=32^o)$~~~ &
~~~${\cal B}$\,\, ($\theta=58^o$)~~~\\
\hline\hline

$  B^{+}\, \to K^{+}~a_1^0$ &  1.4&2.8 \\
\hline

$  B^{+}\, \to K^0~a_1^+$ &  30~~~~(2.0)&75~~~~(5.2)\\
\hline

$  B^{0}\, \to K^0~a_1^0$ &  0.5&2.5 \\
\hline

$  B^{0}\, \to K^{+}~a_1^-$ &  1.6 &4.1 \\
\hline\hline

$  B^{+}\, \to K^{+}~b_1^0$ &  2.6&0.07 \\
\hline

$  B^{+}\, \to K^{0}~b_1^+$ &  4.4~~~~(3.0)&5.0~~~~(0.3) \\
\hline

$  B^{0}\, \to K^{0}~b_1^0$ &  4.1&0.4 \\
\hline

$  B^{0}\, \to K^{+}~b_1^-$ &  2.4&0.2 \\

\hline
\end{tabular}
\end{center}
\end{footnotesize}
\end{table}

In fact we are aware that in some cases the assumptions we make
might be flawed. For example it is known that naive factorization
gives a small contribution to the $B^{0}\, \to \rho^{-} K^+$
channel. The experimental result $\dd {\cal B}(B^{0}\, \to \rho^{-}
K^+)=(9.9^{+1.6}_{-1.5})\times 10^{-6}$ \cite{Group(HFAG):2005rb} is
larger by one order of magnitude than theoretical predictions based
on naive factorization and reasonable models for the form factors
\cite{Ali:1998eb}, which is due to large cancellations between the
penguin contributions appearing in $W_i$. An enhancement with
respect to naive factorization can be due to various reasons. For
example one can mention  ${\cal O}(\alpha_s)$ corrections to the
matrix elements.  Moreover long-distance non-factorizable
contributions, that are power suppressed, such as the charming
penguin contributions \cite{Colangelo:1989gi,Isola:2001ar} are
expected to play a role \cite{Isola:2003fh}, as well as other power
corrections terms in QCD factorization \cite{Zhu:2004}. Finally
including final state interactions requires both perturbative
corrections at leading power, as well as power corrections. The
phenomenology due these effects has been  studied in detail in
\cite{Beneke:2000ry,Beneke:2003zv}. Due to these uncertainties the
results in Table \ref{tab:2} should be interpreted more as tests of
the factorization model than as absolute predictions and are based
on the expectations that, large as they can be, long distance
effects, e.g. those described by final state interactions, cancel
out in the ratios. In any case, to increase our confidence in the
method, we use a different approach to get predictions for these
channels, i.e we consider the ratio of ${\cal B}(B\to {\cal A}_1K)$
to ${\cal B}(B\to \pi K)$. In this case factorization predicts

\bea\frac{{\cal B}(B^+\to K^0 {\cal A}_1^+)_{\rm fact.}}{{\cal
B}(B^+\to K^{0} \pi^+)_{\rm fact.}}\,& =&\frac{{\cal B}(B^0\to K^+
{\cal A}_1^-)_{\rm fact.}}{{\cal B}(B^0\to K^{+} \p^-)_{\rm
fact.}}\,\simeq \,\frac{4}{m^2_B}\,\left(\frac {\dd V_0^{B\to {\cal
A}_1}
(m^2_{K}) } {\dd F_0^ {B\to\pi} (m^2_{K}) }\right)^2\frac{q_{{\cal A}_1}^3}{q_{\p}}\label{pham1}\,,\\
&&\cr \frac{{\cal B}(B^+\to K^+ {\cal A}_1^0)_{\rm fact.}}{{\cal
B}(B^+\to K^{+} \pi^0)_{\rm fact.}}\,&
\simeq&\frac{4}{m^2_B}\,\left|\frac{\frac {\dd V_0^{B\to {\cal A}_1}
(m^2_{K}) } {\dd F_0^ {B\to\pi} (m^2_{K})
}\,+\,\frac{W_3}{W_2}\frac{f_{{\cal A}_1}}{f_K}\,\frac {\dd
F_1^{B\to K} (m^2_{{\cal A}_1}) } {\dd F_0^ {B\to\pi} (m^2_{K})
}}{1\,+\,\frac{W_3}{W_2}\frac{f_{\p}}{f_K}\,\frac {\dd F_0^{B\to K}
(m^2_{\p}) } {\dd F_0^ {B\to\pi} (m^2_{K}) }}\right|^2\frac{q_{{\cal
A}_1}^3}{q_{\p}}\ \label{pham2}\,, \\ &&\cr \frac{{\cal B}(B^0\to
K^0 {\cal A}_1^0)_{\rm fact.}}{{\cal B}(B^0\to K^{0} \pi^0)_{\rm
fact.}}\,& \simeq&\frac{4}{m^2_B}\,\left|\frac{\frac {\dd V_0^{B\to
{\cal A}_1} (m^2_{K}) } {\dd F_0^ {B\to\pi} (m^2_{K})
}\,+\,\frac{W_3}{W_1}\frac{f_{{\cal A}_1}}{f_K}\,\frac {\dd
F_1^{B\to K} (m^2_{{\cal A}_1}) } {\dd F_0^ {B\to\pi} (m^2_{K})
}}{1\,+\,\frac{W_3}{W_1}\frac{f_{\p}}{f_K}\,\frac {\dd F_0^{B\to K}
(m^2_{\p}) } {\dd F_0^ {B\to\pi} (m^2_{K}) }}\right|^2\frac{q_{{\cal
A}_1}^3}{q_{\p}}\label{pham3}\ ; \eea the parameters $W_j$ are in
the  Table of the Appendix A, while for $\frac { V_0^{B\to {\cal
A}_1} (m^2_{K}) } { F_0^ {B\to\pi} (m^2_{K}) }$ we use \be\frac {
V_0^{B\to {\cal A}_1} (m^2_{K}) } { F_0^ {B\to\pi} (m^2_{K})
}\approx\frac { V_0^{B\to {\cal A}_1} (0) } { F_0^ {B\to\pi} (0)
}=\frac { V_0^{B\to {\cal A}_1} (0) } { A_0^ {B\to\r} (0) }\frac {
A_0^{B\to\r} (0) } { F_0^ {B\to\pi} (0) }\,.\ee  We can now compute
again the entries of Table \ref{tab:2} using these formulae and the
experimental {\it BRs} for $B\to K\pi$. The interesting fact is that
in general  we obtain results that differ few percent at most from
those found using the ratios to the decay channels $B\to K\r$. The
only exception is given by the channels $B^{+}\, \to K^0~a_1^+$
  and $B^{+}\, \to K^0~b_1^+$; for them we present both predictions.

Some interesting predictions can be read from  Table \ref{tab:2}.
For $\q=32^o$, for all the decay channels, with the exception of
$a_1^0\,K^0$, we predict {\it BRs} of the order of $10^{-5}$. This
holds also for the channel $K^0a_1^+$, as the prediction based on
$K^0\pi^+$ is more reliable. For $\q=58^o$ we have {\it BRs} of
similar sizes only for $B\to a_1 K$.
 Summing up one can say that
nonleptonic $B$ decays with a kaon and a light non-strange
axial-vector meson in the final state represent interesting decay
channels with generally large {\it BRs}.

\section{$B\to {\cal A}_1 \pi$\label{sec:5}} In this section we consider the
decays $B\to a_1 \pi\ ,~b_1\pi$. To start with, we consider the
channel with at least one neutral particle in final state. We get
the following results for the ratios $\dd\frac{{\cal B}(B\to {\cal
A}_1\, \pi)}{{\cal B}(B\to \rho\, \pi)}$:\bea
\frac{{\cal B}(B^{+}\to {\cal A}_1^0\, \pi^+)_{\rm fact.}}{{\cal B}(B^{+}\to
\rho^0\, \pi^+)_{\rm fact.}} &= &\left(\frac{q_{{\cal
A}_1}}{q_{\rho}}\right)^3\,\left|\frac{\dd\frac{V_0^{B\to {\cal
A}_1}(m_\p^2)}{\dd A_0^{B\to\r}(m_\p^2)}\,w_1\,+w_2\,\frac{\dd
f_{{\cal A}_1}}{f_\pi}\frac{\dd F_1^{B\to \pi}(m^2_{{\cal
A}_1})}{\dd A_0^{B\to\r}(m^2_\pi)}}{w_5\,+\,w_6\,\frac{\dd
f_\r}{f_\pi}\dd\frac{\dd F_1^{B\to \pi}(m^2_\r)}{\dd
A_0^{B\to\r}(m^2_\pi)}}\right|^2\ ,\\ && \cr &&\cr
\frac{{\cal B}(B^{+}\to {\cal A}_1^+\, \pi^0)_{\rm fact.}}{{\cal B}(B^{+}\to
\rho^+\, \pi^0)_{\rm fact.}} &= &\left(\frac{q_{{\cal
A}_1}}{q_{\rho}}\right)^3\,\left|\frac{\dd\frac{V_0^{B\to {\cal
A}_1}(m_\p^2)}{\dd A_0^{B\to\r}(m_\p^2)}\,w_3\,+w_4\,\frac{\dd
f_{{\cal A}_1}}{f_\pi}\frac{\dd F_1^{B\to \pi}(m^2_{{\cal
A}_1})}{A_0^{B\to\r}(m^2_\pi)}}{w_7\,+\,w_3\,\frac{\dd
f_\r}{f_\pi}\frac{\dd F_1^{B\to \pi}(m^2_\r)}{\dd
A_0^{B\to\r}(m^2_\pi)}}\right|^2\ ,\\ && \cr &&\cr
\frac{{\cal B}(B^{0}\to {\cal A}_1^0\, \pi^0)_{\rm fact.}}{{\cal B}(B^{0}\to
\rho^0\, \pi^0)_{\rm fact.}} &= &\left(\frac{q_{{\cal
A}_1}}{q_{\rho}}\right)^3\,\left|\dd\frac{\frac{\dd V_0^{B\to {\cal
A}_1}(m_\p^2)}{\dd A_0^{B\to\r}(m_\p^2)}\,w_4\,+w_2\,\frac{\dd
f_{{\cal A}_1}}{f_\pi}\frac{\dd F_1^{B\to \pi}(m^2_{{\cal
A}_1})}{A_0^{B\to\r}(m^2_\pi)}}{w_7\,+\,w_6\,\frac{\dd
f_\r}{f_\pi}\frac{\dd {F_1^{B\to \pi}(m^2_\r)}}{\dd
A_0^{B\to\r}(m^2_\pi)}}\right|^2\label{a1bis}\,,
\eea

\begin{table}[pht!]
\caption{\label{tab:3} {\small  Theoretical branching ratios  for
$B$ decays into one nonstrange axial-vector meson and a pion for two
different values of the mixing angle. Units are $10^{-6}$.}}
\begin{footnotesize}
\begin{center}
\begin{tabular}{|c|c|c|c|}
\hline ~~~~~~{\rm Process}~~~~~~  & ~~~${\cal B}\,\,
(\theta=32^o)$~~~ &
~~~${\cal B}$\,\, ($\theta=58^o$)~~~&~~~~~~~~~Exp.~~~~~~~~~\\

\hline\hline

$  B^{+}\, \to \pi^{+}~a_1^0$ &  3.9&8.8& $<900$ \cite{Eidelman:2004wy}\\
\hline

$  B^{+}\, \to \pi^{0}~a_1^+$ &  4.8&10.6 &$<1700$ \cite{Eidelman:2004wy}\\
\hline

$  B^{0}\, \to \pi^{0}~a_1^0$ &  1.1&1.7&$<1100$ \cite{Eidelman:2004wy} \\
\hline

\begin{tabular}{c}$B^{0}\, \to \pi^+~a_1^-$\\
$  B^{0}\, \to \pi^{-}~a_1^+$\end{tabular}&
\begin{tabular}{c}4.7\\
11.1\end{tabular}&\begin{tabular}{c}11.8\\
12.3\end{tabular}&$40.9\pm 7.6$ \cite{Abe:2005rf,Aubert:2006dd}\\
\hline\hline

$  B^{+}\, \to \pi^{+}~b_1^0$ &  4.5&0.4 &==\\
\hline

$  B^{+}\, \to \pi^{0}~b_1^+$ &  4.8&0.5&== \\
\hline

$  B^{0}\, \to \pi^{0}~b_1^0$ & 0.5&0.01 &==\\
\hline

$  B^{0}\, \to \pi^{+}~b_1^-$ &  6.9&0.7&== \\ \hline
$  B^{0}\, \to \pi^{-}~b_1^+$ &  $\approx 0$& $\approx 0$&== \\
\hline
\end{tabular}
\end{center}
\end{footnotesize}
\end{table}

We use the ratios in Table \ref{tab:fff} and the experimental {\it
BRs} for  $B\to \rho \pi$ as given by the HFAG group
\cite{Group(HFAG):2005rb}. The results are reported in Table
\ref{tab:3}.

Let us now consider the channels having only charged mesons in the
final state. In order to use the same method employed in the
previous sections we would need the {\it BRs} from the decays
$B^{0}\to \r^+\p^-$ and  $B^{0}\to \r^-\p^+$. Only their sum: ${\cal
B}(B^{0}\to \r^\pm\p^\mp)= {\cal B}(B^{0}\to \r^+\p^-)+ {\cal
B}(B^{0}\to \r^-\p^+)$ is at the moment known ${\cal B}(B^{0}\to
\r^\pm\p^\mp)=(24.0\pm2.5)\times 10^{-6}$ \cite{Group(HFAG):2005rb},
therefore we consider the following ratios (${\cal A
}_1=\,a_1,\,b_1$): \bea \frac{{\cal B}(B^0\to\p^+{\cal A}_1^-)_{\rm
fact.}}{{\cal B}(B^0\to\p^\pm\r^\mp)_{\rm
fact.}}&=&\left(\frac{q_{{\cal
A}_1}}{q_{\rho}}\right)^3\frac{\left|\frac{w_1}{w_5}\frac{\dd
V_0^{B\to {\cal A}_1}(m_\p^2)}{\dd A_0^{B\to\r}(m_\p^2)}\right|^2}
{1+\left|\frac{\dd w_3}{\dd w_5}\frac{\dd f_\r}{\dd f_\p}\frac{\dd
F_1^{B\to\p}(m^2_\r)}{\dd A_0^{B\to\r}(m^2_\p)} \right|^2}~~,
\\
&&\cr &&\cr\frac{{\cal B}(B^0\to\p^-{\cal A}_1^+)_{\rm fact.}}{{\cal
B}(B^0\to\p^\pm\r^\mp)_{\rm fact.}}&=&\left(\frac{\dd q_{{\cal
A}_1}}{q_{\rho}}\right)^3 \frac{\left|\frac{\dd
 w_3}{\dd
 w_5}\frac{\dd
 f_{{\cal A}_1}}{\dd
 f_\p}\frac{\dd
 F_1^{B\to\p}(m^2_{{\cal A}_1})} {\dd
 A_0^{B\to\r}(m^2_\p)}\right|^2}
{1+\left|\frac{\dd
 w_3}{\dd
 w_5}\frac{\dd
 f_\r}{\dd
 f_\p}\frac{\dd
 F_1^{B\to\p}(m^2_\r)}{\dd
 A_0^{B\to\r}(m^2_\p)}\right|^2}~~,\eea
where the parameters $w_k$ are defined in the Appendix A. In this
way we can complete the inputs of Table \ref{tab:3}. There is an
independent analysis, given by H\"ocker et al. \cite{hoecker},
which extracts from  $B^{0}\to \r^\pm\p^\mp$ the values of the
single channels with the result ${\cal B}(B^{0}\to
\r^+\p^-)=(15.3^{+3.7}_{-3.3})\times 10^{-6}$ and ${\cal
B}(B^{0}\to \r^-\p^+)=(14.5^{+4.1}_{-3.6})\times 10^{-6}$.  Using
these values one would  estimate the {\it BRs} of $B^{0}\to {\cal
A}_1^+\, \pi^-$ and $B^{0}\to {\cal A}_1^-\, \pi^+$ with results
$\sim 20\%$ greater than those in Table \ref{tab:3}, i.e within
our estimated theoretical error. A greater confidence can be
obtained using a slightly different approach. One might note that
the ratio $\frac{{\cal B}(B^{0}\to {\cal A}_1^-\, \pi^+)}{{\cal
B}(B^{0}\to \p^-\, \pi^+)}$ is independent of the  Wilson
coefficient and CKM matrix elements. Using as an experimental
input ${\cal B}(B^{0}\to \p^-\, \pi^+)=(4.5\pm0.4)\times 10^{-6}$
\cite{Group(HFAG):2005rb}, one gets values for ${\cal B}(B^{0}\to
{\cal A}_1^-\, \pi^+)$ in agreement with Table \ref{tab:3} within
$\sim10\%$. We note that the prediction for $B^{0}\, \to
\pi^{\mp}~a_1^\pm$ is  somewhat smaller than the result from the
Belle\cite{Abe:2005rf} and BABAR Collaborations
\cite{Aubert:2006dd}, although the value $58^o$ for the mixing
angle offers a better agreement. This is an indication that this
value of the angle is to be preferred. If subsequent analyses
would lead to prefer the solution $\theta=32^o$, this would mean
either a failure of some of our assumptions or that there are
non-resonant effects, not included in the theoretical analysis,
and implicitly taken into account in the data. This might happen
because, for non-resonant diagrams, some particles in the final
state might fall in the same kinematical range as the $a_1$ state,
with an effect  similar to what discussed for $B\to 3\p$ in Ref.
\cite{Deandrea:2000tf}.
\section{Conclusions\label{sec:6}}
In conclusions we have presented predictions for the nonleptonic
$B-$meson decay channel with one axial-vector meson in the final
state. We have used uniquely
 experimental data, e.g. the decay rates for
 $B\to K^*\p\,,K\p\,,\,\r\p,\,\r K$ and, as a theoretical input, the assumption of
 naive factorization. Our results may provide a useful benchmark for
  the future searches of
 the decay channels $B\to K_1\pi$, $B\to a_1 K$, $B\to b_1 K$,
 $B\to a_1\pi\,$, $B\to b_1\pi$ that might be investigated by the
 experimental collaborations.
\appendix
\section{}
In this appendix we list the values of the Wilson coefficients and
the CKM matrix elements we have used in the main text.
\par\noindent
Wilson coefficients (using the results of \cite{Buras:1998ra} for
$\Lambda^{(5)}_{\bar{MS}}=225$ MeV in the HV scheme) and current
quark masses: \bea\{a_1,\ a_2\}=\{1.029,\ 0.140\} ~,~~~
\{a_3,\cdots a_{10}\}&=&\{33,\ -246,\ -10,\ -300,\ 2,\ 4.8,\ -93,\
-12\}\times 10^{-4}\ ,\cr \{m_u,\,m_d,\,m_s\,,m_b\}&=&
\{4,\,8,\,150\,,4600\}~{\text MeV} \ .\eea CKM matrix elements:
\bea
V_{ud}&=&0.97\,,~~~~~~~V_{us}=0.22\,,~~~~~~~~~~~~V_{ub}=0.0018-0.0032\,i\,,~~~\cr
V_{td}&=&0.0074-0.0031\,i\,,~~~V_{ts}=-0.04-0.00072\,i\,,~~~V_{tb}\simeq
1\,. \eea In the text we use some combinations of Wilson
coefficients and CKM matrix elements as reported in Table
\ref{wil}. One may note the correct treatment of isospin
invariance \cite{Pham} in these results.

\renewcommand{\arraystretch}{1.3}
\begin{table}[ht!]
\caption{\small Parameters used in the
paper.\label{wil}}\begin{center}
\begin{tabular}{|c|c|}\hline
{~ Coefficient ~}&Formula\\
\hline

$W_1$&$V^*_{tb}V_{ts}
\left(a_4-\frac12a_{10}+\frac{2m_K^2}{(m_b-m_d)(m_d+m_s)}(a_6-\frac12a_8)
\right)$
\\
$W_2$ & $V^*_{ub}V_{us}a_1-V^*_{tb}V_{ts}
\left(a_4+a_{10}+\frac{2m_K^2}{(m_b-m_u)(m_u+m_s)}(a_6+a_8)\right)$
\\
$W_3$ & $V^*_{ub}V_{us}a_2-V^*_{tb}V_{ts} \frac32(a_9-a_7)$
\\
$W_4$ & $V^*_{tb}V_{ts} \left(a_4-\frac12a_{10}\right)$
\\
$W_5$ & $V^*_{ub}V_{us}a_1-V^*_{tb}V_{ts} \left(a_4+a_{10}\right)$
\\
$W_6$&$V^*_{ub}V_{us}a_1-V^*_{tb}V_{ts}
\left(a_4+a_{10}-\frac{2m_K^2}{(m_b+m_u)(m_u+m_s)}(a_6+a_8)\right)$
\\
$W_7$ & $V^*_{ub}V_{us}a_2-V^*_{tb}V_{ts} \frac32(a_7+a_9)$
\\
$W_8$&$V^*_{tb}V_{ts}
\left(a_4-\frac12a_{10}-\frac{2m_K^2}{(m_b+m_d)(m_d+m_s)}(a_6-\frac12a_8)
\right)$
\\
\hline

$w_1$&$V^*_{ub}V_{ud}a_1-V^*_{tb}V_{td}
\left(a_4+a_{10}+\frac{2m_\pi^2}{(m_b-m_u)(m_u+m_d)}(a_6+a_8)\right)$
\\
$w_2$&$V^*_{ub}V_{ud}a_2-V^*_{tb}V_{td}
\left(-a_4+\frac12a_{10}-\frac32(a_7-a_9)\right)$
\\
$w_3$&$V^*_{ub}V_{ud}a_1-V^*_{tb}V_{td} \left(a_4+a_{10}\right)$
\\
$w_4$&$V^*_{ub}V_{ud}a_2-V^*_{tb}V_{td}
\left(-a_4+\frac12a_{10}-\frac32(a_7-a_9)+\frac{2m_\pi^2}{(m_b-m_d)(m_u+m_d)}(a_6-\frac12a_8)\right)$
\\
$w_5$&$V^*_{ub}V_{ud}a_1-V^*_{tb}V_{td}
\left(a_4+a_{10}-\frac{2m_\pi^2}{(m_b+m_u)(m_u+m_d)}(a_6+a_8)\right)$
\\

$w_6$&$V^*_{ub}V_{ud}a_2-V^*_{tb}V_{td}
\left(-a_4+\frac12a_{10}+\frac32(a_7+a_9)\right)$
\\
$w_7$&~~$V^*_{ub}V_{ud}a_2-V^*_{tb}V_{td}
\left(-a_4+\frac12a_{10}+\frac32(a_9-a_7)-\frac{2m_\pi^2}{(m_b+m_d)(m_u+m_d)}(-a_6+\frac12a_8)\right)$~~
\\
\hline
\end{tabular}
\end{center}
\end{table}

We have used the following definitions for the form factors. If
$|V\rangle$ is a vector meson state ($\rho,\,K^*$) and $|{\cal
A}\rangle$  an axial-vector meson state (i.e one of the states
$K_{1A}\,,K_{1B}\,,a_1\,,b_1$) we use\bea <{V}(\epsilon,p')|V^\mu
-A^\mu|B(p)>& =&\, -\, i (m_B+m_{V}) \epsilon^{*\mu}
 A_1 (q^2) + i \frac{(\epsilon^* \cdot q)}{m_B+m_{V}} (p+p')^\mu A_2
(q^2)\nn\\
& +& i (\epsilon^* \cdot q)\, \frac{2 m_{V}}{q^2} q^\mu\left[ A_3
(q^2)- A_0 (q^2)\right]\,+\,
 \frac {2 V(q^2)}
{m_B+m_{V}} \epsilon^{\mu \nu \alpha \beta}\epsilon^*_{\nu}
p_{\alpha} p'_{\beta}\label{a44}\\ <{\cal A}(\epsilon,p')|V^\mu
-A^\mu|B(p)>& =&\, +\, i (m_B+m_{\cal A}) \epsilon^{*\mu}
 V_1 (q^2) - i \frac{(\epsilon^* \cdot q)}{m_B+m_{\cal A}} (p+p')^\mu V_2
(q^2)\nn\\
& -& i (\epsilon^* \cdot q)\, \frac{2 m_{\cal A}}{q^2} q^\mu\left[
V_3 (q^2)- V_0 (q^2)\right]\,-\,
 \frac {2 A(q^2)}
{m_B+m_{\cal A}} \epsilon^{\mu \nu \alpha \beta}\epsilon^*_{\nu}
p_{\alpha} p'_{\beta} \label{v1}\ . \eea In these equations \be
A_3(q^2) =
  \frac{m_{V}-m_B}{2 m_{V}} A_2(q^2) + \frac {m_{V}+m_B} {2m_{V}} A_1(q^2)
 ~,~~~~
 V_3 (q^2) =\frac {m_{\cal A}-m_B}{2 m_{\cal A}} \,
V_2(q^2)\, + \,\frac {m_{\cal A}+m_B} {2m_{\cal A}} \, V_1(q^2)
\label{v2}\ee with $V_3(0)=V_0(0)$ and $A_3(0)=A_0(0)$,

If $P\,,P^{\prime}$ are pseudoscalar mesons, we have used \be
\langle P^\prime(p^\prime)|V_\mu|P(p)\rangle = F_1(q^2)
\left[(p_\mu+p^\prime_\mu)-\frac{m_P^2-m_{P^\prime}^2}{q^2}\,
q_\mu\right]+ F_0(q^2)\frac{m_P^2-m_{P^\prime}^2}{q^2}\, q_\mu\
.\ee We do not make assumptions on the $q^2$ behavior of the
$F_1^{B\to\pi}$ form factor as we only need its value at $q^2=0$.
Finally we have used the following definitions for the leptonic
decay constants
\be \label{leptonic} \langle 0\, |A_\mu|\, P(p)\rangle = i\, f_P\,
p_\mu ~,~~~~~~~\langle V(\varepsilon,p)|V_\mu|\,0\,\rangle =
 f_V\, m_V\, \varepsilon^\ast_\mu~,~~~~~~\langle {\cal
A}(\varepsilon,p)|A_\mu|\,0\,\rangle =
 f_{\cal A}\, m_{\cal A}\, \varepsilon^\ast_\mu~,
\ee with the following numerical values
$(f_{\pi^+},f_K,f_{\rho^+},f_{K^*})=(132,161,210,210)$ MeV, and,
from $\tau$ decays, $(f_{K_1(1270)},f_{K_1(1400)})=(171,126)$ MeV
\cite{Nardulli:2005fn}.

For the determination of the analogous constants for the $a_1$ and
$b_1$ nonstrange axial-vector mesons one has to take into account
that  the strange axial-vector mesons $K_1(1270)$ and $K_1(1400)$
are the result of the mixing of $^3P_1$ and $^1P_1$ states. Denoting
by $K_{1A}$ and $K_{1B}$ the $^3P_1$ and $^1P_1$ states of $K_1$ one
has \be \label{mix}
 K_1(1270)=K_{1A} \sin\theta+K_{1B}\cos\theta,
 ~,~~~
 K_1(1400)=K_{1A} \cos\theta-K_{1B}\sin\theta\ .
 \ee $K_{1B}$
belongs to the same  nonet as the states  $b_1(1235)$, $h_1(1170)$
and $h_1(1380)$; $K_{1A}$, $a_1(1260)$, $f_1(1285)$ and $f_1(1400)$
are also in one nonet. The mixing angle $\theta$ and the masses of
the $K_1$ states  have been determined in \cite{Nardulli:2005fn}
(but see also \cite{Suzuki:1993yc}, \cite{Cheng:2004yj})  up to a
twofold ambiguity\bea {\rm Sol. [a]}:\hskip1cm\theta&=&32^o\
,\hskip.5cm (m_{K_{1B}},\,m_{K_{1A}})=(1310,\ 1367)\ {\rm MeV}\ ,\cr
{\rm Sol. [b]}:\hskip1cm\theta&=&58^o\ ,\hskip.5cm
(m_{K_{1B}},\,m_{K_{1A}})=(1367,\ 1310)\ {\rm MeV}\ .\eea Using this
result and $SU(3)$ symmetry one gets \cite{Nardulli:2005fn}: \bea
{\rm Sol. [a]}\hskip1cm(\theta&=&32^o):\hskip.5cm
(f_{b_{1}},\,f_{a_{1}})=(74,\ 215)\ {\rm MeV}\ ,\cr {\rm Sol.
[b]}\hskip1cm(\theta&=&58^o):\hskip.5cm
(f_{b_{1}},\,f_{a_{1}})=(-28,\ 223)\ {\rm MeV}\ .\eea The two values
for $f_{b_1}$ given above refer to the strange partner in the octet
and come from $SU(3)$ violations \cite{Suzuki:1993yc}, the non
strange partners having zero coupling.

We have also used the matrix element describing radiative
transitions:\bea\langle K_1(p^\prime,\,\epsilon)|{\bar
s}\sigma_{\mu\nu}(1+\gamma_5) q^\nu b|
B(p)\rangle\,&=&\,i\epsilon_{\mu\nu\rho\sigma}\epsilon^{*\,\nu}p^\rho
p^{\prime\,\sigma}2\,T_1(q^2)\,+\,[\epsilon_\mu^{*}(m^2_B-m^2_{K_1})\,-\,(\epsilon^*\cdot
q)(p+p^\prime)_\mu]\,T_2(q^2)\cr&+&[(\epsilon^{*}\cdot
q)q_\mu\,-\,\frac{q^2}{m^2_B-m^2_{K_1}}(p+p^\prime)_\mu]\,T_3(q^2) \
, \label{t1} \eea with $T_1(0)=T_2(0)$ ($T_3$ does not contribute to
the radiative decay). For $B\to K^*$ an analogous formula can be
written.  From experiment one has \cite{Nardulli:2005fn}, \be y=
\frac{T_1^{B\to K_{1}(1270)}(0)}{T_1^{B\to K^{*}}(0)}\approx
1.06~,~~~ y^\prime= \frac{T_1^{B\to K_{1}(1400)}(0)}{T_1^{B\to
K^{*}}(0)}~\approx\bigg\{
\begin{array}{c}0.14~~~(\theta=32^o)\\0.35~~~~(\theta=58^o)\end{array}
\ .\label{y}\ee

 \vspace{0.2truecm} \textbf{Acknowledgements}

 We thank P. Santorelli
 for most useful discussions, F. Palombo and J. Zupan for discussions on the
BABAR data and J. Smith for valuable correspondence.

\vskip.1cm


\end{document}